\documentclass[twocolumn,twoside]{IEEEtran}
\usepackage{times}
\usepackage{amsmath}  
\usepackage{amssymb}  
\usepackage{mathrsfs} 
\usepackage{theorem}  
\usepackage{cite}     
\usepackage{comment}  
\usepackage{upref}
\usepackage{amsfonts}
\usepackage{verbatim}
\usepackage[dvipsnames,usenames]{color}
\usepackage{enumerate}
\usepackage{graphicx}
\usepackage{subfigure}
\usepackage{latexsym}
\usepackage{caption}
\usepackage{algorithm}
\usepackage{psfrag}
\usepackage[normalem]{ulem}
\usepackage{setspace}

\usepackage{color}
\usepackage{times,color}
\usepackage[usenames,dvipsnames,svgnames,table]{xcolor}

\usepackage[top=0.53in, bottom=0.53in, left=0.6in, right=0.6in]{geometry}

\usepackage{algpseudocode}

\usepackage{tikz}
\usetikzlibrary{shapes,arrows,positioning,decorations.pathreplacing,calc}

\newcommand{\ve}[1]{\mathbf{#1}}
\newcommand{\err}[1]{\ve{\underline{#1}}}

\newcommand{\code}{\ttfamily\bfseries}

\newcommand{\be}[1]{\begin{equation}\label{#1}}
\newcommand{\ee}{\end{equation}}

\newcommand{\bc}{\begin{center}}
\newcommand{\ec}{\end{center}}

\newcommand{\qed}{\hfill$\Box$\\[1ex]}

\newcommand{\cC}{{\cal C}}

\newcommand{\cH}{{\cal H}}

\newcommand{\mS}{\mathcal{S}}

\newcommand{\bfu}{{\boldsymbol u}}
\newcommand{\bfv}{{\boldsymbol v}}
\newcommand{\bfw}{{\boldsymbol w}}
\newcommand{\bfx}{{\boldsymbol x}}
\newcommand{\bfy}{{\boldsymbol y}}

\newcommand{\LM}{{L_M}}
\newcommand{\sml}{{\Sigma_{M,L}}}

\newcommand{\mle}{{\mathsf{le}}} 

\renewcommand{\le}{\leqslant}
\renewcommand{\leq}{\leqslant}
\renewcommand{\ge}{\geqslant}
\renewcommand{\geq}{\geqslant}

\newcommand{\Cref}[1]{Co\-rol\-la\-ry\,\ref{#1}}

\theoremstyle{plain} \theorembodyfont{\normalfont\slshape}

\newtheorem{thm}{Theorem$\!$}
\newenvironment{theorem}{\begin{thm}\hspace*{-1ex}{\bf.}}{\end{thm}}

\newtheorem{prop}[thm]{Proposition$\!$}

\newtheorem{lem}[thm]{Lemma$\!$}
\newenvironment{lemma}{\begin{lem}\hspace*{-1ex}{\bf.}}{\end{lem}}

\newtheorem{cor}[thm]{Corollary$\!$}
\newenvironment{corollary}{\begin{cor}\hspace*{-1ex}{\bf.}}{\end{cor}}

\newtheorem{const}[thm]{Construction$\!$}

\newtheorem{cl}[thm]{Claim$\!$}

\newtheorem{conject}[thm]{Conjecture$\!$}

\newtheorem{defi}[thm]{Definition$\!$}
\newenvironment{definition}{\begin{defi}\hspace*{-1ex}{\bf.}}{\end{defi}}

\theorembodyfont{\normalfont}

\newtheorem{exam}{Example$\!$}

\newtheorem{remrk}{Remark$\!$}

\definecolor{Codecolor}{named}{White}  

\newcommand{\Copen}{\mbox{\{\kern-5.50pt\{}}
\newcommand{\Cclose}{\mbox{\}\kern-5.50pt\}}}
\newcommand{\Cslash}{\mbox{$\backslash\kern-6.02pt\backslash$}}

\newcommand*{\script}[1]{\mathcal{#1}}

\begin{document}
\title{\textbf{Clustering-Correcting Codes}}

\author{
\textbf{Tal Shinkar},\!\IEEEauthorrefmark{1}
\textbf{Eitan Yaakobi},\!\IEEEauthorrefmark{1}
\textbf{Andreas Lenz},\!\IEEEauthorrefmark{2}
and \textbf{Antonia Wachter-Zeh},\!\IEEEauthorrefmark{2}
   
\IEEEauthorblockA{\IEEEauthorrefmark{1}Department of Computer Science, Technion --- Israel Institute of Technology, Haifa, 3200009 Israel \\ }
\IEEEauthorblockA{\IEEEauthorrefmark{2}Institute for Communications Engineering, Technical University of Munich, Munich 80333, Germany\\ }
{\code  \small \{tal.s,yaakobi\}@cs.technion.ac.il},\,
{\code \small andreas.lenz@mytum.de, antonia.wachter-zeh@tum.de}\vspace{-5ex}}
\maketitle

\thispagestyle{empty}
\pagestyle{empty}

\begin{abstract}
A new family of codes, called \emph{clustering-correcting codes}, is presented in this paper. This family of codes is motivated by the special structure of data that is stored in DNA-based storage systems. The data stored in these systems has the form of unordered sequences, also called \emph{strands}, and every strand is synthesized thousands to millions of times, where some of these copies are read back during sequencing. Due to the unordered structure of the strands, an important task in the decoding process is to place them in their correct order. This is usually accomplished by allocating a part of the strand for an index. However, in the presence of errors in the index field, important information on the order of the strands may be lost.

Clustering-correcting codes ensure that if the distance between the index fields of two strands is small, then there will be a large distance between their data fields. It is shown how this property enables to place the strands together in their correct clusters even in the presence of errors. We present lower and upper bounds on the size of clustering-correcting codes and an explicit construction of these codes which uses only a single bit of redundancy.

\end{abstract}\vspace{-2.5ex}
\section{Introduction}

Progress in synthesis and sequencing technologies have paved the way for the development of a non-volatile data storage based on DNA molecules. The first large-scale experiments that demonstrated the potential of in vitro DNA storage were reported by Church et al. who recovered 643 KB of data \cite{CGK12} and Goldman et al. who accomplished the same task for a 739 KB message \cite{GBCDLSB13}. However, in both of these works the data was not recovered successfully due to the lack of using the appropriate coding solutions to correct errors. Since then, several more groups have demonstrated the ability to successfully store data of large scale using DNA molecules; see e.g.~\cite{BLCCSS16,BGHCTIPC16,EZ17,Oetal17,YTMZM15}. Other works developed coding solutions which are specifically targeted to correct the special types of errors inside DNA-based storage systems~\cite{J17,LSWY18,Retal17,YTMZM15,LY18,YK15,Y16}.			


A DNA storage system consists of three steps. 
First, the strands containing the encoded data are synthesized. They are then stored inside a storage container, and finally a DNA sequencer reads back the strands. 
The encoding and decoding are two external processes to the system that convert the data to DNA strands and back. The structure of a DNA storage system is  different from all other existing storage systems. Since the strands are stored unordered, it is unclear what part of the data each strand represents, even if no error occurred. For more details we refer the reader to~\cite{HMG18,LSWY18} and referencers therein.

Storing DNA strands in a way that will allow to reconstruct them back in the right order is an important task. The common solution to address this problem is to use indices, that are stored as part of the strand. Each strand is prefixed with some nucleotides that indicate the strand's location, with respect to all other strands. Although using indices is a  simple solution it has several drawbacks. One of them is that in case of an error within the index, important information on the strand's location is lost as well as the ability to place it in the correct position between the other strands. In this paper a new coding scheme, called \emph{clustering-correcting codes}, is presented which enables to combat errors with minimal redundancy. 

In DNA storage systems, every strand is synthesized thousands of times (or even millions) and thus more than a single copy of each strand is read back upon sequencing. Thus, the first task based upon the sequencer's input is to partition all the reads into clusters such that all read strands at each cluster are copies of the same information strand. A possible solution is to use the indices in order to identify the strands and cluster them together, but in the presence of errors, this may result with mis-clustered strands which can cause errors in the recovered data. Hence, finding codes and algorithms for the clustering process is an important challenge. A naive solution is to add redundancy to the index part in order to correct potential errors in the index~\cite{CKW19}. However, this will incur an unavoidable reduction in the storage rate of the DNA storage system. We will show in this paper how clustering-correcting codes can enable one to cluster all strands in the right clusters even with the presence of errors in the indices (see Fig.~\ref{fig:channel}), while the redundancy is minimized. In fact, for a large range of parameters this can be done with only a single bit of redundancy for all the strands together.

The rest of the paper is organized as follows. In Section~\ref{sec:def}, the family of clustering-correcting codes is presented as well as other useful definitions that will be used throughout the paper. In Section~\ref{sec:bounds}, we present explicit and asymptotic lower and upper bounds on the size of clustering-correcting codes. In Section~\ref{sec:alg}, we present an explicit construction of these codes which uses only a single bit of redundancy. Due to the lack of space, some of the proofs are omitted in the paper. 
	 
	\vspace{-2ex} 
\section{Definitions and Preliminaries} \label{sec:def}	

For a positive integer $n$, the set $\{0,1,\ldots,n-1\}$ is denoted by $[n]$. For two vectors $ \bfx, \bfy \in \{ 0, 1 \} ^ n $ of the same length,  we denote the $i$-th symbol of $ \bfx $ by $ x_i $. The subvector of $\bfx$ starting at the $i$-th index of length $\ell$ is denoted by $\bfx_{[i,\ell]} $. 
The Hamming distance between $\bfx$ and $\bfy$ is denoted by $ d_H(\bfx, \bfy)$ and the Hamming weight of $\bfx$ is $w_H(\bfx)$. 
The radius-$r$ ball of a vector $\bfx\in\{0,1\}^ n$ is $B_r(\bfx) = \{\bfy \ | d_H(\bfx,\bfy)\leq r\}$ and its size is denoted by 
$B_n(r) \triangleq \sum_{i=0}^r\binom{n}{i}$. The function $\cH(x)$ for $0\leq x\leq 1$ denotes the binary entropy function, and the inverse function $\cH^{-1}(x)$ for $0\leq x\leq 1$ is defined to return values between 0 and 1/2. 
We study here  the binary case while extensions to the non-binary case are straightforward.

	\begin{figure*}[ht!]
	\begin{center}
		\newcommand{\xwidth}{0cm}
		\newcommand{\indexwidth}{1.25cm}
		\newcommand{\datawidth}{2.04cm}
		\newcommand{\mht}{0.55cm}
		\tikzstyle{block} = [rectangle, draw, minimum height=\mht]
		\begin{tikzpicture}

		\node[block, fill = lightgray!50!white, minimum width=\indexwidth] (index) {$00$};
		\node[block, right= 0pt of index, minimum width=\datawidth] (data) {$110011$};
		
		\node[block, fill = lightgray!50!white, below= .05cm of index,minimum width=\indexwidth] (index2) {$01$};
		\node[block, right= 0pt of index2, minimum width=\datawidth] (data2) {$001000$};
		
		\node[block, fill = lightgray!50!white, below= .05cm of index2,minimum width=\indexwidth] (index3) {$10$};
		\node[block, right= 0pt of index3, minimum width=\datawidth] (data3) {$111100$};
		
		\node[block, fill = lightgray!50!white, below= .05cm of index3,minimum width=\indexwidth] (index4) {$11$};
		\node[block, right= 0pt of index4, minimum width=\datawidth] (data4) {$000111$};

		\node[block, fill = lightgray!50!white, minimum width=\indexwidth] at ($(data.east) + (2.5cm,\mht)$) (Bindex) {$10$};
		\node[block, right= 0pt of Bindex, minimum width=\datawidth] (Bdata) {$111100$};
		
		\node[block, fill = lightgray!50!white, below= .05cm of Bindex,minimum width=\indexwidth] (Bindex2) {$01$};
		\node[block, right= 0pt of Bindex2, minimum width=\datawidth] (Bdata2) {$001000$};
		
		\node[block, fill = lightgray!50!white, below= .05cm of Bindex2,minimum width=\indexwidth] (Bindex3) {$\err{1}0$};
		\node[block, right= 0pt of Bindex3, minimum width=\datawidth] (Bdata3) {$110011$};
		
		\node[block, fill = lightgray!50!white, below= .05cm of Bindex3,minimum width=\indexwidth] (Bindex4) {$00$};
		\node[block, right= 0pt of Bindex4, minimum width=\datawidth] (Bdata4) {$110\err{1}11$};
		
		\node[block, fill = lightgray!50!white, below= .05cm of Bindex4,minimum width=\indexwidth] (Bindex5) {$00$};
		\node[block, right= 0pt of Bindex5, minimum width=\datawidth] (Bdata5) {$110011$};
		
		\node[block, fill = lightgray!50!white, below= .05cm of Bindex5,minimum width=\indexwidth] (Bindex6) {$10$};
		\node[block, right= 0pt of Bindex6, minimum width=\datawidth] (Bdata6) {$11110\err{1}$};

		\node[block, fill = lightgray!50!white, minimum width=\indexwidth] at ($(Bdata.east) + (2.5cm,0)$) (Cindex) {$00$};
		\node[block, right= 0pt of Cindex, minimum width=\datawidth] (Cdata) {$110\err{1}11$};
		
		\node[block, fill = lightgray!50!white, below= .05cm of Cindex,minimum width=\indexwidth] (Cindex2) {$00$};
		\node[block, right= 0pt of Cindex2, minimum width=\datawidth] (Cdata2) {$110011$};
		
		\node[block, fill = lightgray!50!white, below= .15cm of Cindex2,minimum width=\indexwidth] (Cindex3) {$01$};
		\node[block, right= 0pt of Cindex3, minimum width=\datawidth] (Cdata3) {$001000$};
		
		\node[block, fill = lightgray!50!white, below= .15cm of Cindex3,minimum width=\indexwidth] (Cindex4) {$\err{1}0$};
		\node[block, right= 0pt of Cindex4, minimum width=\datawidth] (Cdata4) {$110011$};
		
		\node[block, fill = lightgray!50!white, below= .05cm of Cindex4,minimum width=\indexwidth] (Cindex5) {$10$};
		\node[block, right= 0pt of Cindex5, minimum width=\datawidth] (Cdata5) {$111100$};
		
		\node[block, fill = lightgray!50!white, below= .05cm of Cindex5,minimum width=\indexwidth] (Cindex6) {$10$};
		\node[block, right= 0pt of Cindex6, minimum width=\datawidth] (Cdata6) {$11110\err{1}$};

		\node at ($(index) + (\datawidth/2, .75)$) {$\mS$};
		\node[text width = 1.3cm] at ($(index) + (\datawidth+1.5cm, .8)$) {Draw \& Perturb};
		\node at ($(Bindex) + (\datawidth/2, .75)$) {$G$};

		\draw[->] (data3.east) -- (Bindex.west);
		\draw[->] (data2.east) -- (Bindex2.west);
		\draw[->] (data.east) -- (Bindex3.west);
		\draw[->] (data.east) -- (Bindex4.west);
		\draw[->] (data.east) -- (Bindex5.west);
		\draw[->] (data3.east) -- (Bindex6.west);
		
		\draw[->] ($(Bdata3.south east) + (0.1,0.025)$) -- node[above] {Cluster} node[below] {indices} ($(Cindex3.south west) + (-0.1,0.125)$);
		
		\draw[->] (Cdata4.east) to[out=0,in=0] node [right, text width = 2.5cm] {Identify outlier and move to correct cluster} (Cdata2.south east);
		
		\draw[dashed] ($(Cindex.north west) + (-0.05,0.05)$) rectangle ($(Cdata2.south east) + (0.05,-0.05)$);
		\draw[dashed] ($(Cindex3.north west) + (-0.05,0.05)$) rectangle ($(Cdata3.south east) + (0.05,-0.05)$);
		\draw[dashed] ($(Cindex4.north west) + (-0.05,0.05)$) rectangle ($(Cdata6.south east) + (0.05,-0.05)$);

		\end{tikzpicture}
		\caption{Exemplary realization of the DNA channel model. A set $\mS$ of $M=4$ strands is stored and $N = 6$ strands are drawn with errors (highlighted in bold). The strands are clustered according to their indices. The outlier can be identified as it has large distance w.r.t. all other strands in the cluster and be put into the correct cluster.}
		\label{fig:channel}
		\end{center}
		\vspace{-.6cm}
	\end{figure*}
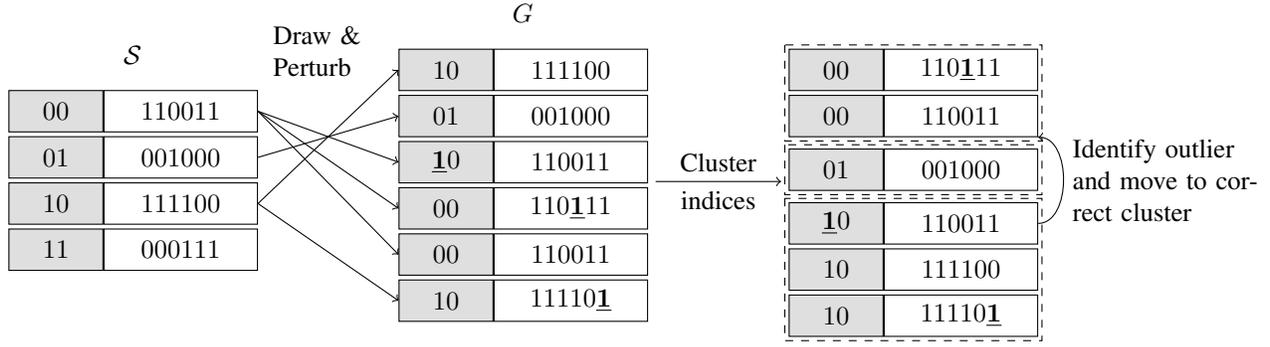

Assume $ M $ strands are stored in a DNA-based storage system where the size of every strand is $L$. We will assume that $M=2^{\beta L}$ for some $0<\beta < 1$ and for simplicity, it is assumed that $M$ is a power of $ 2 $. For any integer $i \in[M]$,  its binary representation of length $ \log (M) $ is denoted by $ \mathsf{ind}_i $. Every length-$L$ strand $s$ that will be stored in the system is of the form $s=(\mathsf{ind},\bfu)$, where $\mathsf{ind}$ is the length-$ \log (M)$ \emph{index field} of the strand (the binary representation of a number between $0$ and $M-1$) and $\bfu$ is the \emph{data field} of $L-\log (M)$ bits that are used to store the information or the redundancy of an error-correcting code. Every stored message will have $M$ strands of this form and the space of all possible messages that can be stored in the DNA storage system is 
$$ \hspace{-0.5ex}\script{X}_ {M\hspace{-0.3ex}, L} \hspace{-0.5ex} = \hspace{-0.5ex} \{ \{ (\mathsf{ind}_0,\hspace{-0.3ex} \bfu_0), \ldots, (\mathsf{ind}_ {M - 1},\hspace{-0.3ex} \bfu_ {M - 1}) \} |  \bfu_j \hspace{-0.5ex}\in\hspace{-0.5ex} \{0, 1\}^ {L-\log (M)} \}. $$
Clearly, $ | \script{X} _ {M, L} | = 2 ^  {M (L-\log (M))} $. Under this setup, a code $ \script{C}$ will be a subset of $ \script{X} _ {M, L}$, where each codeword $\mS$ of $\script{C}$ is a subset of the form $\mS = \{(\mathsf{ind}_0, \bfu_0), \ldots, (\mathsf{ind}_ {M - 1}, \bfu_ {M - 1}) \}$. For shorthand, 
the term $L-\log (M)$ will be abbreviated by $\LM$.

When a set $\mS = \{(\mathsf{ind}_0, \bfu_0), \ldots, (\mathsf{ind}_ {M - 1}, \bfu_ {M - 1}) \}$ is synthesized, each of its strands $(\mathsf{ind}_i,\bfu_i)$, which are called the \emph{input strands}, has thousands to millions of copies and during the sequencing process a subset of these copies is read. Hence, the sequencer's output is another set $G$ of some $N$ strands, called the \emph{output strands}, where $N$ is significantly larger than $M$. Each output strand in the set $G$ is a copy of one of the input strands in $\mS$, however with some potential errors. A DNA-based storage system is called a \emph{$(\tau,\rho)$-DNA system} if it satisfies the following property: If the output strand $(\mathsf{ind}',\bfu')\in G$ is a noisy copy of the input strand $(\mathsf{ind},\bfu)\in S$, then $d_H(\mathsf{ind},\mathsf{ind}')\leq \tau$ and $d_H(\bfu,\bfu')\leq \rho$. That is, the index field has at most $\tau$ Hamming errors while the data field has at most $\rho$ Hamming errors. We consider in this work only substitution errors while extensions for deletions, insertions, and more generally the edit distance will be analyzed in the full version of the paper.

Since the set $G$ contains several noisy copies of each input strand in $\mS$, the first task in the decoding process is to partition the set of all $N$ output strands into $M$ cluster sets, such that the output strands in every cluster are noisy copies of the same input strand. Since every strand contains an index, the simplest way to operate this task is by partitioning the output strands into $M$ sets based upon the index field in every output strand. This process will indeed be successful if there are no errors in the index field of every output strand, however other solutions are necessary since the error rates in DNA storage systems are not negligible~\cite{HMG18}. Another approach to cluster the strands is based upon the distances between every pair of output strands, as was studied in~\cite{Retal17}. However, this approach suffers extremely high computational complexity. 

In this work, we take a hybrid solution of these two approaches. First, we cluster the output strands based on the indices in the output strands. Then, we scan for output strands which were mis-clustered, that is, were placed in the wrong cluster. This is accomplished by checking the distances between the output strands in every cluster in order to either remove output strands that were incorrectly placed in a cluster due to errors in their index or move them to their correct cluster set. Since we compute the distances only between pairs of strands that were placed in the same cluster (and not between all pairs of strands), this step will result in a significantly lower complexity compared to the solution from~\cite{Retal17}. However, in order to succeed in this new approach we need the strands stored in the set $\mS$ to satisfy several constraints. These constraints will be met by the family of \emph{clustering-correcting codes} which are presented in this paper. Another assumption taken in this model, which will be referred to as the \emph{majority assumption}, assumes that in every cluster the majority of the strands have the correct index. Since the number of strands is very large this assumption holds with  high probably if not in all cases.

The main idea to move strands which were misplaced in a cluster due to errors in their index field works as follows. Assume the strand $s_i=(\mathsf{ind}_i, \bfu_i)$ has a noisy copy of the form $s_i'=(\mathsf{ind}'_i, \bfu'_i)$, and let $j$ be such that $s_j = (\mathsf{ind}_j,\bfu_j)$ and $\mathsf{ind}_j = \mathsf{ind}'_i$. We need to make sure that the distance between $\bfu'_i$ and $\bfu_j$ is large enough as 
this will allow to identify that the output strand $s_i'$ is erroneous and therefore does not belong to the cluster of index $\mathsf{ind}_j$; see Fig.~\ref{fig:channel}. We will be interested in either identifying that the output strand $s_i'$ does not belong to this cluster or more than that, place it in its correct cluster. This motivates us to study the following family of codes. \vspace{-2ex}
\begin{definition}
A word $ \mS = \{(\mathsf{ind}_0, \bfu_0), \ldots, (\mathsf{ind}_ {M - 1}, \bfu_ {M - 1}) \}\in\script{X} _ {M,L} $ is said to satisfy the $(e,t)$-\textbf{clustering constraint} if for all $ (\mathsf{ind}_i, \bfu_i), (\mathsf{ind}_j, \bfu_j) \in \mS $ in which $i\neq j$ and $ d_H (\mathsf{ind}_i, \mathsf{ind}_j) \le e $, it holds that $d_H(\bfu_i, \bfu_j) \ge t $. 

A code $ \script{C} \subseteq \script{X} _ {M,L} $ will be called an $(e,t)$-\textbf{clustering-correcting code} (\textbf{CCC})  if every $\mS\in \cC$ satisfies the $(e,t)$-clustering constraint. \vspace{-1.5ex}
\end{definition} 
The \emph{redundancy} of a code $ \script{C} \subseteq \script{X}_{M,L} $ will be defined by \vspace{-1ex}
$$r = M  \LM - \log |\script{C}|.\vspace{-1ex}$$
Our goal in this work is to find $ (e,t)$-CCCs for all $ e $ and $ t $. 
We denote by $A_{M,L}(e,t)$ the size of the largest $(e,t)$-CCC in $ \script{X} _ {M,L} $, and by $r_{M,L}(e,t)$ the optimal redundancy of an $(e,t)$-CCC, so $r_{M,L}(e,t) = M  \LM - \log (A_{M,L}(e,t)).$

The clustering-correcting capabilities of CCCs are proved in the next theorem. We note that as a result of this theorem, if the number of errors is not too large, it is already possible to place every output strand in its correct cluster.\vspace{-1.5ex}
\begin{theorem}
For fixed integers $M,L,e,t$, let $\cC$ be an $(e,t)$-CCC. Assume that a set $\mS\in \cC$ is stored in a $(\tau,\rho)$-DNA system. The following properties hold:
\begin{enumerate}
\item \label{item:ident} If $ \tau \le e $ and $ 4 \rho < t $ then every output strand can be detected to be placed in a wrong cluster. 
\item \label{item:correct}  If $ \tau \le e/2  $ and $ 4 \rho < t $ then every output strand can be placed in its correct cluster.\vspace{-1ex}
\end{enumerate}
\end{theorem}
\begin{IEEEproof}
We prove only the first statement while the proof of second one is similar. 
Let $ (\mathsf{ind}_i', \bfu_i') $ be a noisy copy of the strand $ (\mathsf{ind}_i, \bfu_i) $. Since the data is stored in a $(\tau, \rho) $-DNA system, it holds that $ d_H(\mathsf{ind}_i, \mathsf{ind}_i') \le \tau $, and therefore $ d_H(\mathsf{ind}_i, \mathsf{ind}_i') \le e $. Also $ d_H(\bfu_i, \bfu_i') \le \rho $. Let $j\in[M]$ be such that $\mathsf{ind}_j=\mathsf{ind}_i'$. From the fact that $ S \in \cC $ we  derive that $ d_H(\bfu_i, \bfu_j) \ge t > 4 \rho $, and thus 
$ d_H(\bfu_j, \bfu_i') \geq d_H(\bfu_i, \bfu_j) - d_H(\bfu_i, \bfu_i') >3 \rho. $ Let $ (\mathsf{ind}_j, \bfu_j') $ be a noisy copy strand of $ (\mathsf{ind}_j, \bfu_j) $, that is, errors might occur in the data field but not in the index field. So, $ d_H(\bfu_j, \bfu_j') \le \rho $ which yields that $ d_H(\bfu_i', \bfu_j') > 2 \rho $. On the other hand, the distance between the data fields of the two strands that belong to the same cluster is at most $ 2\rho $. That is, under the majority assumption,
a mis-clustered strand will have a distance of more than $ 2 \rho $ from the majority of the strands in the cluster, and so it can be dropped instead of being mis-clustered.

\end{IEEEproof}

\vspace{-3ex}

\section{Bounds}\label{sec:bounds}
Upper and lower bounds on $A_{M,L}(e,t)$ are presented.  Let $ D_n(d) $ be the size of the largest {$\textmd{length-}n$} error-correcting code $C\subseteq \{0,1\}^n$ and minimum Hamming distance $d$. For the rest of the paper, let $B_1 = B_{\log(M)}(e) -1, B_2= B_{\LM}(t-1)$, $\mle = \log(\exp(1)) \approx 1.44 $, and it is also assumed that $\beta<1/2$.
\vspace{-1.5ex}
\begin{theorem}\label{th:low}
For all $M,L, e,$ and $t$ it holds that\vspace{-1.5ex}
$$ A_{M,L}(e,t) \ge  2 ^ {M  \LM} \left(1-\frac {B_1  B_2} {2^{\LM}} \right)^{M-D} \vspace{-1ex}$$
and hence \vspace{-2ex}
$$r_{M,L}(e,t) \leq \frac {\mle \cdot(M - D)  B_1  B_2} {2^{\LM} - B_1   B_2},$$
where $D= D_{\log(M)}(e+1)$.
\end{theorem}\vspace{-1ex}
\begin{IEEEproof}
In order to verify the lower bound, we construct an $(e,t)$-CCC $\cC$ that will yield a lower bound on $ A_{M,L}(e,t)$. 
Let $C_1\subseteq \{0,1\}^{\log(M)}$ be a length-$\log(M)$ code with minimum Hamming distance $e+1$ of size $D$. Each codeword  $\mS =\{(\mathsf{ind}_0, \bfu_0), \ldots, (\mathsf{ind}_ {M - 1},\bfu_ {M - 1})\}\in\cC$ is constructed in two steps. First, we choose the data field of strands with indices that belong to the code $C_1$, that is, all strands of the form $(\mathsf{ind},\bfu)$ such that $\mathsf{ind}\in C_1$. There are $2 ^ {\LM}$ options for each strand and thus $(2 ^ {\LM})^{D}$ options for the first step. Since the Hamming distance between all pairs of indices of these strands is at least $e+1$, their data fields can be chosen independently. 

For the rest of the strands we assume the worst case. That is, for each strand left, all of its neighbors are chosen and their radius-$(t-1)$ balls are mutually disjoint. Thus, there are at least $$2 ^ {\LM} - (B_{\log (M)}(e) -1)\cdot B_{\LM}(t-1) = 2 ^ {\LM} - B_1 B_2 $$ options to choose the data field of each remaining strand. In conclusion, there are 
$ 2^{\LM D} \left( 2^{\LM} -  B_1   B_2\right)^{M - D} $
options for choosing a valid set $\mS \in \cC $, and hence \vspace{-1ex}
$$ A_{M,L}(e,t) \ge  2 ^ {M  \LM} \left(1 - \frac {B_1   B_2} {2^{\LM}} \right)^{M - D}.\vspace{-1ex}$$

We can also deduce an upper bound on the redundancy
$$r_{M,L}(e, t) \leq \frac { \mle \cdot (M - D)  B_1 B_2} {2^{\LM} - B_1   B_2},$$
where here the inequality $-\log(1-x)\leq \mle\cdot \frac{x}{1-x}$ for all $0<x<1$ is used.
\end{IEEEproof}

The next corollary follows directly from Theorem~\ref{th:low}.
\begin{corollary}\label{cor:tbound}
If $t\hspace{-0.5ex}\leq \hspace{-0.5ex}\LM \hspace{-0.5ex}\cH^{-1}\hspace{-0.7ex}\left( \frac{1-2\beta}{1-\beta} \hspace{-0.5ex}-\hspace{-0.5ex}\frac{\log(\beta L)}{(1-\beta)L}\right)$ then $r_{M,L}(1,t) \hspace{-0.5ex}\leq\hspace{-0.5ex} 1$.
\end{corollary}
\begin{IEEEproof}
For $ e = 1$, $D_{\log(M)}(2) = M/2 $. This is achieved by selecting all indices $ \mathsf{ind}_i $ such that $ w_H(\mathsf{ind}_i) $ is even (or odd). 
Hence, from Theorem~\ref{th:low} it holds that
$$ r_{M,L}(1,t) \leq \frac {\mle\cdot M \log(M)  B_2} {2^{\LM + 1} - 2 \log(M) B_2}.$$
Hence, $r_{M,L}(1,t)\leq 1$ if 
$B_2 \leq \frac{2^{\LM + 1}}{\log(M)(\mle \cdot M+2)}.$
According to Lemma 4.7 in~\cite{R05}, $B_2\leq 2^{\LM \cH\left(\frac{t-1}{\LM}\right)}$ and hence it is enough to require that  
$2^{\LM \cH\left(\frac{t-1}{\LM}\right)} \leq \frac{2^{\LM + 1}}{\log(M)(\mle \cdot M+2)},$
i.e., 
\begin{small}
\begin{align*}
\LM \cH\left(\frac{t-1}{\LM}\right) & \leq \LM -\log(M) -\log\log(M)& \\
&  \leq   \LM + 1 - \log(\mle \cdot M+2) -\log\log(M). & 
\end{align*}
\end{small}
For $M=2^{\beta L}$, this holds for all $t\leq \LM \cH^{-1}\left( \frac{1-2\beta}{1-\beta} -\frac{\log(\beta L)}{(1-\beta)L} \right)$. 
\end{IEEEproof}

A similar upper bound on $A_{M,L}(e,t)$ is presented next.\vspace{-1ex}
\begin{theorem}\label{th:up}
For all $M,L, e$ and $t$ it holds that\vspace{-1ex}
$$A_{M,L}(e,t) \le 2 ^ {M \LM} \left( 1 - \frac{B_2} {2^{\LM}}\right)^{M - 1},\vspace{-1ex}$$
and therefore 
$r_{M,L}(e,t) \geq \frac {\mle \cdot(M - 1) \cdot B_2} {2^{\LM}}.$ Furthermore, 
if ${t\geq\LM\cdot \cH^{-1}\left( \frac{1-2\beta}{1-\beta} +\frac{\log(\LM)}{\LM}\right)+1}$ then $r_{M,L}(1,t) \geq 1$.
\end{theorem}

From Corollary~\ref{cor:tbound} and Theorem~\ref{th:up} we get that for $\LM$ large enough and $t\approx \LM\cdot \cH^{-1}\left( \frac{1-2\beta}{1-\beta}\right)$, we get that 
$r_{M,L}(1,t) \approx 1$. An asymptotic improvement to the upper bound from Theorem~\ref{th:up}
for $e=1$ which matches the lower bound from Theorem~\ref{th:low} is proved in the next theorem. \vspace{-1ex}
\begin{theorem}
For $\LM$ large enough, if $\frac{t}{\LM} < \frac{1}{2}  \cH^{-1}\left(  \frac{1-2\beta}{1-\beta}\right) -\epsilon$, for some fixed $\epsilon>0$, then it holds that\vspace{-1ex}
$$A_{M,L}(1,t) \leq 2 ^ {M  \LM}\left(1 -\frac{\log(M)  B_2} {2^{\LM}} \right)^{M/2} (1+\delta),$$
for $\delta$ small enough, and hence \vspace{-1ex}
$$r_{M,L}(1,t) \geq \frac {M  \log(M) B_2} {2^{\LM+1} - 2\log(M)  B_2} - \log(1+\delta).$$
\end{theorem}
\begin{IEEEproof}
Let $\cC$ be a $(1,t)$-CCC of maximal size $A_{M,L}(1,t)$. 
For every set $\mS = \{(\mathsf{ind}_0, \bfu_0), \ldots, (\mathsf{ind} _ {M - 1}, \bfu_ {M - 1}) \} \in \cC$, let
$$\mS_{\textmd{even}} = (\bfu_i)_{i: w_H(\mathsf{ind}_i) \textmd{ is even}} \in (\{0,1\}^{\LM})^{M/2} \triangleq \sml$$ be the \emph{vector} projection of $\mS$ to the data fields of the strands with indices of even weight 
and let  $I_{\textmd{even}} \triangleq\{i \ | \ w_H(\mathsf{ind}_i) \textmd{ is even}\}$.

The sets of the strands in the code $\cC$ are partitioned according to their projection on the indices with even weight. More specifically, for every $\bfv \in \sml$, let $\cC_{\bfv}$ be the subcode of $\cC$,
$$\cC_{\bfv} = \{  \mS\in \cC \ | \  \mS_{\textmd{even}} = \bfv\}, \vspace{-1ex}$$
so it holds that $\cC = \bigcup_{\bfv \in \sml}  \cC_{\bfv}$. 

A vector $\bfv = (\bfv_i)_{i\in I_{\textmd{even}}}  \in \sml$ is \emph{good} if for all $i,j\in I_{\textmd{even}}$ such that $d_H(\mathsf{ind}_i,\mathsf{ind}_j)=2$ it holds that $B_{t-1}(\bfv_i)\cap B_{t-1}(\bfv_j) =\emptyset$, and otherwise it is \emph{bad}. Denote by $X_{\textmd{good}},X_{\textmd{bad}}$ the number of good, bad vectors in $\sml$, respectively. 
If a vector $\bfv \in \sml$ is bad, then there are at least two indices $i,j\in I_{\textmd{even}}$ such that $d_H(\mathsf{ind}_i,\mathsf{ind}_j)=2$ and $B_{t-1}(\bfv_i)\cap B_{t-1}(\bfv_j) \neq\emptyset $, i.e., $d_H(\bfv_i,\bfv_j)\leq 2t-2$. Hence, we get that 
\begin{align*}
X_{\textmd{bad}} & \leq  M (\log(M))^2 B_3 \cdot 2^{\LM(\frac{M}{2}-1)}, & 
\end{align*}
where $B_3 = B_{\LM}(2t-2)$. 

Consider the size of $\cC_{\bfv}$ when $\bfv$ is good. For every $\mS\in\cC_{\bfv}$, we only need to assign the data fields for strands of odd weight index. Since $\bfv$ is a good vector, for every index of odd weight, the radius-$(t-1)$ balls of all of its neighbor strands are mutually disjoint so there are exactly
$$2 ^ {\LM} - \log (M)\cdot B_2 $$ options to choose the data field of the $i$-th strand. For every bad vector $\bfv \in \sml$, it is enough to take the loose bound in which $|\cC_{\bfv}|\leq \left( 2 ^ {\LM}  \right)^{\frac{M}{2}}$.
In conclusion we get that 

\vspace{-2ex}
\begin{small}
\begin{align*}
|\cC| & = \Big|\bigcup_{\bfv \in \sml}  \cC_{\bfv} \Big| = \Big|\bigcup_{\bfv \in \sml : \bfv \textmd{ is good} }  \cC_{\bfv} \Big| + \Big|\bigcup_{\bfv \in \sml : \bfv \textmd{ is bad} }  \cC_{\bfv} \Big| & \\
& \leq  X_{\textmd{good}} \left(2 ^ {\LM} - \log (M) B_2\right)^{\frac{M}{2}} + X_{\textmd{bad}} \left( 2 ^ {\LM}  \right)^{\frac{M}{2}} & \\
& \leq 2^{\frac{M\LM}{2}} \left(2 ^ {\LM} - \log (M) B_2\right)^{\frac{M}{2}} + X_{\textmd{bad}} 2 ^ {\frac{M\LM}{2}} & \\
& \leq 2^{M \LM} \left(1 - \frac{\log (M) B_2}{2 ^ {\LM}}\right)^{\frac{M}{2}} + \frac{M (\log(M))^2 B_32^{M\LM}}{2^{\LM}} & \\
& = 2^{M \LM} \left(1 - \frac{\log (M) B_2}{2 ^ {\LM}}\right)^{\hspace{-0.5ex}\frac{M}{2}} \hspace{-1ex}\left( \hspace{-0.5ex} 1 + \frac{M (\log(M))^2 B_3}{2^{\LM}\cdot \left(1 - \frac{\log (M)\cdot B_2}{2 ^ {\LM}}\right)^{\frac{M}{2}}}  \hspace{-0.5ex}\right). &  \vspace{-2ex}
\end{align*}\end{small}
According to $-\log(1-x)\leq \mle\cdot \frac{x}{1-x}$ for all $0<x<1$ we get
$$ \left(1 - \frac{\log (M) B_2}{2 ^ {\LM}}\right)^{\frac{M}{2}} \geq 2^{-\frac{ \mle\cdot \log(M) B_2(M/2)}{2^{\LM}-\log(M) B_2}}. $$
We use again the inequality $B_2\leq 2^{\LM \cH\left(\frac{t-1}{\LM}\right)}$ and  $B_3\leq 2^{\LM \cH\left(\frac{2(t-1)}{\LM}\right)}$, while for $\frac{t}{\LM} < \frac{1}{2}  \cH^{-1}\left(  \frac{1-2\beta}{1-\beta}\right) -\epsilon$ it holds 
\begin{small}
\begin{align*}
& \lim_{\LM\rightarrow\infty}\frac{\mle\cdot\log(M) B_2(M/2)}{2^{\LM}-\log(M) B_2} \leq \lim_{\LM\rightarrow\infty}\frac{\mle\cdot\log(M) 2^{\LM \cH\left(\frac{t-1}{\LM}\right)}2^{\beta L-1} }{2^{\LM}-\log(M) 2^{\LM \cH\left(\frac{t-1}{\LM}\right)}}  & \\
& = \lim_{\LM\rightarrow\infty}\frac{\mle\cdot\log(M) 2^{\LM \cH\left(\frac{t-1}{\LM}\right)}2^{\beta L-1} }{2^{\LM}}  = 0,&
\end{align*}
\end{small}
and 
\begin{align*}
& \lim_{\LM\rightarrow\infty}  \frac{M (\log(M))^2 B_3}{2^{\LM} } \leq \lim_{\LM\rightarrow\infty}  \frac{M (\log(M))^2 2^{\LM \cH\left(\frac{2(t-1)}{\LM}\right)}}{2^{\LM} } & \\
& = \lim_{L\rightarrow\infty}  \frac{2^{\beta L} (\beta L)^2 2^{(1-2\beta-\epsilon')L} }{2^{(1-\beta)L} } = \lim_{L\rightarrow\infty}  \frac{(\beta L)^2}{2^{\epsilon' L} } = 0,& 
\end{align*}
for some $\epsilon'>0$. Thus, $  \lim_{L\rightarrow\infty} \frac{M L^{2t}}{2^{\LM}\cdot \left(1 - \frac{\log (M)\cdot B_2}{2 ^ {\LM}}\right)^{\frac{M}{2}}}  =0,$
which confirms the theorem's statements. \vspace{-1ex}

\end{IEEEproof}

\section{A Construction of CCCs}\label{sec:alg}
In this section we propose a construction of CCCs. It is shown that with a single bit of redundancy it is possible to construct CCCs for relatively large values of $t$. 

The algorithm will use the following functions:
\begin{itemize}
	\item The function $ w_\ell(S, t) $ is defined over a set of vectors $S$ and a positive integer $t$ and outputs a vector $ \bfw \in  \{0,1\}^\ell$ which satisfies the following condition. For all $\bfv \in S$,  $d_H(\bfw, \bfv_{[\log (M), \ell]})  \ge t $. The value of $\ell$ will be determined later as a function of $e$, $t$, and $M$.
	\item The function $ \Delta_1(\mathsf{ind}_i, \mathsf{ind}_j) $ encodes the difference between the two indices $i$ and $j$ of Hamming distance at most $e$ using $e\lceil\log(\log(M))\rceil$ bits which mark the positions where the indices $\mathsf{ind}_i, \mathsf{ind}_j$ differ. 
	\item The function $\Delta_2(\bfu_i, \bfu_j)$ encodes the difference between the two data fields $\bfu_i, \bfu_j\in \{0,1\}^{L_M}$ of Hamming distance at most $t-1$ using $(t-1)\log(\LM)$ bits which mark the positions where they differ. 	
\end{itemize}

The input to the algorithm is $ M $ vectors $ \bfv_0, \ldots, \bfv_{M-1} $. All vectors are of length $\LM$ bits, except for $ \bfv_{M-1} $ which has length of $ \LM  - 1$ bits. The idea behind the presented algorithm is to find all pairs of vectors that do not satisfy the clustering constraint, and correct them in a way that they satisfy the constraint and yet the original data can be uniquely recovered. A bit is added to $ \bfv_{M-1} $, hence, the code has a single bit of redundancy, to mark whether some vectors were altered by the algorithm.

For $i\in[M]$, the notation $S(e,i)$ in the algorithm will be used as a shortcut to the set $\{\bfu_j \ | \ d_H(\mathsf{ind}_i,\mathsf{ind}_j)\leq e\}$ of data fields corresponding to indices $\mathsf{ind}_j$ of Hamming distance at most $e$ from $\mathsf{ind}_i$. At any iteration of the while loop, when the $i$-th strand is corrected, the function $w_\ell(S(e,i), t)$ will be used to update the data field $ \bfu_i$ such that it does not violate the constraint and yet can be decoded. 
In order to make room for the vector generated by the function $w_\ell(S(e,i), t)$, we will encode $ \bfu_i $ based on its similarity to one of its close neighbors $\bfu_j$. These modifications are encoded together as a repelling vector of length $ len = \ell + \log \log (M) + (t - 1) \log (\LM)$, when $e=1$. \vspace{-2ex}
\begin{algorithm}[hptb]
	\caption{ $ (1,t)-CCC $ Construction}\label{alg:cons}
	\setstretch{1.2}
	\begin{algorithmic}[1]
		\Require \hspace{-0.4ex}$ M $ vectors $\hspace{-0.2ex} \bfv_0, \ldots, \bfv_{M - 1} $ such that $ \bfv_0, \ldots, \bfv_{M - 2} \in \{ 0, 1 \} ^ {\LM} $ 
			and $ \bfv_{M - 1} \in \{ 0, 1 \} ^ {\LM - 1}  $
		\Ensure a codeword $ \mS = \{ (\mathsf{ind}_0, \bfu_0), \ldots, (\mathsf{ind}_{M-1}, \bfu_{M-1}) \} $
		\State $ \forall i \in \{ 0, \ldots, M - 2 \} : \bfu_i = \bfv_i, \bfu_{ M - 1} = (\bfv_{ M - 1}, 0) $
		\State $ p = M - 1 $
		\State\label{state:bad}$B = \{ (i, j) \ | \ i < j, d_H(\mathsf{ind}_i, \mathsf{ind}_j) \le 1 \wedge d_H(\bfu_i, \bfu_j) < t \} $
		\While { $ B \neq \emptyset $, let $ (i, j) \in B $ }
		\State\label{step_mark} $ (\bfu_p)_{\LM - 1} = 1 $
		\State $ (\bfu_p)_{[0, \log (M)]} = \mathsf{ind}_i $
		\State $ p = i $
		\State \label{state:bb} $ repl = (w_\ell(S(1,i), t), \Delta_1(\mathsf{ind}_i, \mathsf{ind}_j), \Delta_2(\bfu_i, \bfu_j)) $
		\State $ (\bfu_i)_{[\log (M), len]} = repl $
		\State \label{state:update} $ B = \{ (i, j) \ | \ (i, j) \in B \wedge d_H(\bfu_i, \bfu_j) < t \} $
		\EndWhile
		\State $ (\bfu_p)_{\LM - 1} = 0 $
		\State $ (\bfu_p)_{[0, \log (M)]} = (\bfv_{M - 1})_{[0, \log (M)]}$
	\end{algorithmic}
\end{algorithm}\vspace{-3ex}
\begin{theorem}\label{tm:cons}
For any input vectors $ \bfv_0 \ldots, \bfv_{M-1} $ , Algorithm \ref{alg:cons} returns a valid $ (1, t)$-CCC codeword for any $ t $ that satisfies:\vspace{-1ex}
$$ \ell + \log \log (M) + (t - 1) \log (\LM) + 1 \leq \LM - \log (M).\vspace{-1ex}$$
Furthermore, it is possible to decode the  vectors $ \bfv_0 \ldots, \bfv_{M-1}$.\vspace{-1ex}
\end{theorem}

\begin{IEEEproof}
After the initializing steps, 
in Step~\ref{state:bad} the algorithm gathers the indices of all pairs of strands that both their index and data fields are too close to each other, hence, violating the constraint. The algorithm iterates over the set $ B $, handling one pair at a time. In Step~\ref{state:update}, this set is updated and the algorithm stops when the set $ B $ is empty, i.e., there are no bad pairs and so $  \{ (\mathsf{ind}_0, \bfu_0), \ldots, (\mathsf{ind}_{M-1}, \bfu_{M-1}) \} $ satisfies the constraint. 

On each iteration of the while loop, the algorithm takes a pair of strands, say of indices $i$ and $j$, where $i<j$, which violates the constraint and changes the data field in the $i$-th strand. First, the flag at the end of the previous strand is changed to $ 1 $ (Step~\ref{step_mark}). This denotes that it is not the last strand in the decoding chain. In Step~\ref{state:bb}, the algorithm calculates the vector $ \bfw = w_\ell(S(1,i), t) $ and embeds it in the data field of the ${i\textmd{-th}}$ strand. The vector $\bfw$ satisfies that for all $\bfu \in S(1,i)$, $d_H(\bfw, \bfu_{[\log (M), \ell]})  \ge t $.  Then, for all $ \bfu \in S(1,i)$ we have that $d_H(({\bfu_i})_{[\log (M), \ell]}, \bfu_{[\log (M), \ell]}) \ge t$ and lastly for all $\bfu \in S(1,i)$, $d_H(\bfu_i, \bfu) \ge t $. Therefore the $i$-th and the $j$-th strands satisfy the constraint and thus do not belong to the set $B$ when it is updated in Step~\ref{state:update}. In fact any bad pair of indices which includes the $i$-th strand will be removed as well from the set $B$. Furthermore, since the $i$-th strand has been updated in such a way that it satisfies the constraint with respect to all of its neighbors, and the $p$-th strand in this iteration already satisfies the constraint according to his last $\LM-\log (M)$ bits, no bad pairs have been created. That is, the size of the set $B$ decreases in each iteration, and the algorithm terminates. The constraint on $t$ guarantees that the data field is large enough in order to write the information required on each update step of the while loop.


The idea of the decoding process is to track the updates chain of the strands 
and then traverse the chain in the opposite direction while recovering the input vectors. 
\end{IEEEproof}

Algorithm~\ref{alg:cons} can easily be extended to support larger values of $e$. In this case the constraint on $ t $ is
$$ L - 2 \log (M) \ge \ell + \log (B_{\log (M)}(e) \cdot B_{\LM}(t - 1)) + 1 .$$


Next we discuss the function $w_\ell(S, t) $. This function takes a set of vectors $ S $ as input, and outputs a vector $ \bfw \in\{0,1\}^\ell$ such that for all $\bfv \in S_{[\log (M), \ell]} \triangleq \{ \bfv_{[\log (M), \ell]} | \bfv \in S \} $, it holds that 
$ d_H(\bfw, \bfv)  \ge t $. The length $\ell$ of the vector $\bfw$ is determined by the minimum value of $\ell$ for which\vspace{-1ex} $$ 2^{\ell} > |S_{[\log (M), \ell]}| \cdot B_{\ell}(t - 1). $$ That is, the minimum length that allows us to choose a vector that does not fall into any of the radius-$(t-1)$ balls of the vectors in the set $S_{[\log (M), \ell]}$. 
The size of $S_{[\log (M), \ell]}$ is at most $ B_{\log (M)} (e) - 1$, and we denote by $\ell(e,t,M)$ the smallest value of $\ell$ such that 
$ 2^{\ell} > (B_{\log (M)} (e) - 1) \cdot B_{\ell}(t - 1).$
\begin{lemma}\label{lem:bb}
For all $e,t,M$ such that $ t \leq \frac {(\log(M)) ^ e} {e \cdot \log \log (M)}, $ it holds that 
$$\ell(e,t,M) \leq e t \cdot \log\log(M).$$
\end{lemma}

\begin{corollary}
For all      $t\le \frac {\LM - \log (M) - e \log \log (M) + \log (\LM) } { \log(\LM) + e \log\log(M)} $
there exists an explicit construction of an $(e,t)$-CCC using Algorithm~\ref{alg:cons} which uses a single bit of redundancy.
\end{corollary}
\begin{IEEEproof}
From Lemma~\ref{lem:bb} we can use $ w_\ell(S, t) $ in Algorithm~\ref{alg:cons} with $ \ell = e t \cdot \log \log (M) $. In addition, from Theorem~\ref{tm:cons} the value of $t$ should satisfy 
$$ \ell + \log (B_{\log (M)}(e) \cdot B_{\LM}(t - 1)) + 1 \leq \LM - \log (M).$$
Thus, it is enough to show that 
$$ e  t \cdot \log \log (M) + \log ((\log(M))^e \cdot \LM^{t-1}) \leq \LM - \log (M),$$
while using $B_{\log (M)}(e) \leq (\log(M))^e$ and $B_{\LM}(t - 1) \leq \LM^{t-1}$. 
Hence, it is enough for $t$ to satisfy 
\begin{small}
$$t(\log (\LM) + e \log \log (M)) \leq \LM - \log(M)-e  \log \log (M) + \log(\LM).\vspace{-1ex}$$
\end{small}
\end{IEEEproof}

According to Section~\ref{sec:bounds}, $r_{M,L}(1,t) = 1$ when $t$ is approximately $\LM\cdot \cH^{-1}\left( \frac{1-2\beta}{1-\beta}\right)$.
However, this is not achieved by an explicit construction of such codes. Here, we presented an explicit construction for $e=1$ in which the maximum value of $t$ is roughly $\frac{\LM}{\log(\LM)}\frac{1-2\beta}{1-\beta}$. That is, at most only a factor of $\log(\LM)$ from the theoretical upper bound on $t$.

\vspace{-1ex}


\end{document}